\documentclass[aps,prb,reprint]{revtex4-2}
\usepackage{amsmath,amssymb,graphicx,numprint,times,dcolumn}
\usepackage[usenames,dvipsnames,svgnames,table]{xcolor}
\usepackage[colorlinks=true, urlcolor=blue, linkcolor=blue, citecolor=blue, pdftex]{hyperref}
\usepackage{orcidlink}

\npthousandsep{\,}
\npdecimalsign{.}

\def\<{\langle}
\def\>{\rangle}
\def\dphi{\Delta_{\hat{\phi}}}

\newcolumntype{.}{D{.}{.}{4}}
\newcolumntype{o}{D{.}{.}{7}}
\newcolumntype{q}{D{.}{.}{1}}

\newcommand{\PA}{Physica}
\newcommand{\PAA}{\PA~A}

\newcommand{\JSTAT}{J.~Stat.~Mech.}
\newcommand{\JSP}{J.~Stat.~Phys.}

\newcommand{\PR}{Phys.~Rev.}
\newcommand{\PRL}{Phys.~Rev.~Lett.}

\newcommand{\PRB}{Phys.~Rev. B}

\newcommand{\PRE}{Phys.~Rev. E}
\newcommand{\PRX}{Phys.~Rev. X}

\newcommand{\JHEP}{J.~High Energy Phys.}

\newcommand{\JPCM}{J.~Phys.: Condens.~Matter}
\newcommand{\JPAOLD}{J.~Phys.~A: Math.~Gen.}
\newcommand{\JPA}{J.~Phys.~A: Math.~Theor.}
\newcommand{\physrep}{Phys.~Rep.}

\newcommand{\ZPB}{Z.~Phys.~B}

\newcommand{\NPB}{Nucl.~Phys.~B}

\newcommand{\RMP}{Rev.~Mod.~Phys}

\begin{document}

\title{The ordinary surface universality class of the three-dimensional O($N$) model}
\author{\firstname{Francesco} \surname{Parisen Toldin}\,\orcidlink{0000-0002-1884-9067}}
\email{parisentoldin@physik.rwth-aachen.de}
\affiliation{Institute for Theoretical Solid State Physics, RWTH Aachen University, Otto-Blumenthal-Str. 26, 52074 Aachen, Germany}
\affiliation{JARA-FIT and JARA-CSD, 52056 Aachen, Germany}
\begin{abstract}
  We study the critical behavior at the ordinary surface universality class of the three-dimensional O($N$) model, bounded by a two-dimensional surface.
  Using high-precision Monte Carlo simulations of an improved lattice model, where the leading bulk scaling correction is suppressed, and finite-size scaling analysis of the fourth cumulant of the surface magnetization,
  we obtain precise estimates of the scaling dimension of the surface field operator for $N=2,3,4$.
  We also determine the fixed-point values of two renormalization-group invariant observables, which characterize the finite-size scaling behavior at the ordinary transition.
\end{abstract}

\maketitle

\section{Introduction}
Critical phenomena in the presence of boundaries or, more generally, defects is a rich field of study, which has attracted over the years numerous experimental \cite{Dosch-book} and theoretical \cite{Binder-83,Diehl-86,Pleimling-review} studies.
A general renormalization-group (RG) analysis shows that a bulk universality class (UC), describing the critical behavior in the thermodynamic limit, generically splits into several boundary UCs, leading to a rich phase diagram \cite{Cardy-book}.
Furthermore, critical exponents and other universal quantities at boundaries differ from the bulk ones \cite{Binder-83,Diehl-86}.
The simplest setup realizing this framework consists in a $d-$dimensional critical system bounded by a $(d-1)-$dimensional surface: different surface UCs can be then realized by tuning surface couplings.
Surface UCs are also relevant for the critical Casimir force \cite{FG-78,Krech-94,Krech-99,BDT-00,Gambassi-09,GD-11,MD-18,DD-23}.
Despite being a mature subject, boundary critical phenomena has recently received renewed attention.
The discovery of unexpected boundary exponents in some quantum spin models has sparked numerous investigations \cite{SS-12,ZW-17,DZG-18,WPTW-18,WW-19,JXWX-20,ZDZG-20,WW-20,DZGZ-21,WZG-22,SLL-22,YHSXDZ-22,SL-22,XPXZ-21}.
At the same time, recent progresses in conformal field theory have addressed the problem of boundary and defects in conformally-invariant models \cite{MO-95,LRVR-13,GLMR-15,BGLM-16,LM-17,LMT-18,MRZ-19,KP-20,DHS-20,BDPLVR-20,GGLVV-21,BCFT-review,PKMGM-21}.
Closely related to boundary critical phenomena is the research field of the so-called gapless topological states of matter, and in particular their boundary states \cite{GV-12,BQ-14,CCBCN-15,BMF-15,SPV-17,PSV-18,Verresen-20,VTJP-21,TVV-21}.

In this context, recent advancements have challenged the understanding of the bulk-surface phase diagram of the paradigmatic three-dimensional O($N$) UC \cite{PV-02}, in the presence of a 2D surface.
For an isolated 2D surface, the Mermin-Wagner-Hohenberg theorem \cite{MW-66,*MW-66_erratum,Hohenberg-67,Halperin-19} and its generalizations \cite{Cassi-92,MW-94} forbid long-range order for $N\ge 2$.
While a 2D O(2) model exhibits the Berezinskii-Kosterlitz-Thouless transition, for $N>2$ the 2D O($N$) model is always disordered \cite{PV-02}.
These results are expected to hold also for a surface next to a disordered bulk, for in this case one can imagine to integrate out the bulk degrees of freedom, leading to an effective short-ranged O($N$)-invariant interaction on the surface \cite{Diehl-86,Diehl-thanks}.
For a critical bulk, and for $N>2$, the above considerations may suggest that the surface would not host a phase transition, since the topology of the phase diagram does not dictate it.
In contrast with this argument, in Ref.~\cite{PT-20} we have shown that the surface of a 3D O(3) model exhibits a \textit{special} transition, separating the \textit{ordinary} phase with the \textit{extraordinary} one \footnote{Earlier hint of surface transition were found in Ref.~\cite{DBN-05}.}.
In fact, a recent field-theoretical analysis has pointed out the existence, for a finite range $2\le N\le N_c$, of a new surface UC of the 3D O($N$) model, dubbed  ``extraordinary-log'',
where the two-point function of the order parameter decays as a power of a logarithm \cite{Metlitski-20}.
Its existence and the logarithmic exponent
is determined by some universal RG amplitudes of the so-called \textit{normal} surface UC, which is realized by imposing a boundary symmetry-breaking field.
In Ref.~\cite{PTM-21} we have extracted these amplitudes by means of Monte Carlo (MC) simulations for $N=2,3$.
A comparison with direct simulations of the extraordinary-log UC reveals a good agreement, thus quantitatively confirming the nontrivial relation between the normal and the extraordinary-log UC.
The field-theoretical analysis of the boundary critical behavior has been recently extended to the case of a plane defect in the three-dimensional O($N$) model, where it has been shown that the extraordinary-log phase exists for all $N$ \cite{KM-23}.
The extraordinary-log UC has been investigated in various settings \cite{HDL-21,DZGZ-21,SHL-23}.

While the extraordinary and the special surface UCs in the 3D O($N$) model exist for some values of $N$ only, the ordinary UC is always present;
it can be generically realized on the surface of a critical O($N$) system, without enhancement of the surface interactions.
At the ordinary UC, there is only a single relevant surface operator, corresponding to the order parameter, and its two-point function decays quickly, such that the surface susceptibility is finite \cite{Binder-83,Diehl-86}.
A previous MC determination of the ordinary surface critical exponent for $N=2$, $3$ \cite{DBN-05} displays a small discrepancy with truncated conformal bootstrap (TCB) results \cite{GLMR-15}; the latter is, however, affected by a systematic error, whose magnitude is difficult to estimate \cite{GLMR-15,PKMGM-21}.
For $N=4$, we are only aware of the MC study of the ordinary UC in Ref.~\cite{Deng-06}.

In this Letter we provide a precise numerical determination of the scaling dimension of the surface field operator $\dphi$ at the ordinary UC, for $N=2,3,4$.
To this end, we employ high-statistics MC simulations of an ``improved'' model, where the leading scaling corrections are suppressed, and a finite-size scaling (FSS) analysis of a higher-order cumulant of the surface field.
Our results will provide a benchmark for future studies, and in particular for the conformal bootstrap approach \cite{CB_review}.

\section{Model}
We study the classical $\phi^4$ model on a three-dimensional $L\times L\times L$ lattice, imposing periodic boundary conditions (BC) on two directions, and open BC along the remaining one, thus realizing two surfaces.
The reduced Hamiltonian ${\cal H}$, such that the Gibbs weight is $\exp(-\cal H)$, is
\begin{equation}
    {\cal H} = -\beta\sum_{\< i\ j\>}\vec{\phi}_i\cdot\vec{\phi}_j
    -\beta_{s}\sum_{\< i\ j\>_{s}}\vec{\phi}_i\cdot\vec{\phi}_j
    +\sum_i[\vec{\phi}_i^{\,2}+\lambda(\vec{\phi}_i^{\,2}-1)^2],
  \label{model}
\end{equation}
where $\vec{\phi}_x$ is an $N-$components real field on the lattice site $x$, the first sum extends over the nearest-neighbor pairs where at least one site belongs to the inner bulk, the second sum over the sites on the two surfaces, and the last term is summed over all lattice sites.
In the Hamiltonian (\ref{model}) the coupling constants $\beta$ and $\lambda$ determine the bulk behavior, whereas $\beta_s$ control the surface interactions; here, we consider an identical coupling strength on the two surfaces.

\begin{table}
  \caption{Estimates of the value of $\lambda=\lambda^*$ for which the model (\ref{model}) is improved, for $N=2,3,4$. In the third column we report the value of the coupling constant $\beta=\beta_c$ at the onset of the critical point, for a value of $\lambda$ within the uncertainty interval of $\lambda^*$.}
  \begin{ruledtabular}
    \begin{tabular}{llr}
      $N$ & \multicolumn{1}{c}{$\lambda^*$} & \multicolumn{1}{c}{$(\lambda,\beta_c(\lambda))$} \\
      \hline
      2 & 2.15(5)  \cite{CHPV-06}       & (2.15, \numprint{0.50874988}(6))  \cite{PT-21}\\
      3 & 5.17(11) \cite{Hasenbusch-20} & (5.2, \numprint{0.68798521}(8))   \cite{Hasenbusch-20}\\
      4 & 18.4(9)  \cite{Hasenbusch-21} & (18.5, \numprint{0.91787555}(17)) \cite{Hasenbusch-21}\\
    \end{tabular}
  \end{ruledtabular}
  \label{table.improved}
\end{table}

In the limit $\lambda\rightarrow\infty$, the model reduces to the standard hard spin O($N$) model. In the $(\lambda, \beta)$ plane, the bulk phase diagram exhibits a second-order transition line in the O($N$) UC \cite{CHPRV-02,PV-02,CHPV-06,Hasenbusch-21}.
Along this critical line, for $N\le 4$ there is a specific point $(\lambda^*, \beta_c(\lambda^*))$ where the model is improved, i.e., the leading scaling correction vanishes.
In Table \ref{table.improved} we report the improvement value $\lambda^*$ and the corresponding critical coupling $\beta=\beta_c(\lambda)$ for $N=2,3,4$, as determined in previous studies \cite{CHPV-06,PT-21,Hasenbusch-20,Hasenbusch-21}.
Improved models are a rather useful tool to obtain accurate results in numerical studies of critical phenomena \cite{PV-02}, in particular in the presence of boundaries \cite{Hasenbusch-09b,Hasenbusch-10c,PTD-10,Hasenbusch-11,Hasenbusch-11b,Hasenbusch-12,PTTD-13,PT-13,PTTD-14,PTAW-17,PT-20,PTM-21}, whose presence gives rise to additional corrections to scaling, which cumulate with those originating from bulk irrelevant perturbations.

To realize the ordinary UC, we fix $\beta$ and $\lambda$ to the values reported in the third column of Table \ref{table.improved}, thus tuning the bulk to its critical point, and set $\beta_s=\beta$.
This choice corresponds to the absence of enhancement of the surface interactions, and generically realizes the ordinary UC; further surface phases and a transition may be explored by tuning the surface interaction strength, while keeping the bulk to its critical point \cite{Binder-83,Diehl-86}.
We numerically sample the model by means of MC simulations, combining Metropolis, overrelaxation, and Wolff single-cluster updates \cite{Wolff-89}; details of the simulation algorithm are reported in Ref.~\cite{PT-20}.
To improve the statistics, for every surface observable we perform an average of the values sampled on the two identical surfaces.

\section{Results}
The scaling dimension of the surface field or, equivalently, of the surface field operator can be computed by a FSS analysis of the correlations of the lattice field $\phi$ on the surface.
To this end, the most commonly used quantity is the surface susceptibility $\chi_s$, defined as
\begin{equation}
  \chi_s \equiv \frac{1}{L^2} \<\vec{M}_s\cdot \vec{M}_s \>, \qquad \vec{M}_s\equiv \sum_{i\in S}\vec{\phi}_i.
  \label{chis}
\end{equation}
where the sum in the definition of the surface magnetization $\vec{M}_s$ extends over the sites on one surface.
By a standard FSS analysis \cite{Privman-90}, at the ordinary critical point and in a finite size $\chi_s$ scales as
\begin{equation}
  \chi_s = A L^{2-2\dphi} + B,
  \label{FSSchis}
\end{equation}
where $\dphi$ is the scaling dimension of the relevant O($N$)-vector operator at the surface, $A$ and $B$ are two nonuniversal constants, and we have neglected scaling corrections.
The scaling dimension of the surface operator $\dphi$ is related to that of the surface field $y_{h_1}$ by $\dphi = 2 - y_{h_1}$, and to the critical exponent $\eta_\parallel$ by $\dphi=(1+\eta_\parallel)/2$.
As is well known, at the ordinary UC $\chi_s$ is finite \cite{Binder-83,Diehl-86}, hence the exponent in Eq.~(\ref{FSSchis}) $2-2\dphi<0$, and the scaling behavior of $\chi_s$ is dominated by the nonuniversal background term $B$.
In fact, as shown below, the exponent of the singular part is $2-2\dphi\approx -0.4$; its smallness exacerbates the FSS analysis of $\chi_s$, because one needs to clearly separate the background term from the slow-decaying singular part $\propto L^{-0.4}$.
On top of that, scaling corrections not considered in Eq.~(\ref{FSSchis}) further hinder a precise scaling analysis of $\chi_s$, rendering this observable not suitable for a quantitatively accurate determination of $\dphi$.
To overcome this problem, we have sampled the fourth cumulant $\chi_{4s}$ of the surface magnetization.
It can be defined by considering an external field $\vec{h}_s$ on a single surface and coupled to the surface magnetization $\vec{M}_s$, therefore adding to the reduced Hamiltonian (\ref{model}) a term $-\vec{h}_s\cdot\vec{M}_s$.
The fourth cumulant $\chi_{4s}$ is then defined as
\begin{equation}
  \chi_{4s} \equiv -\frac{1}{L^2}\left.\frac{\partial^4 f}{\left(\partial \vec{h}_s\cdot\partial \vec{h}_s\right)^2}\right|_{\vec{h}_s=0},
  \label{chi4_def}
\end{equation}
where $f \equiv -\ln Z$ is the free energy in units of $k_BT$, with $Z$ the partition function; the factor $1/L^2$ in Eq.~(\ref{chi4_def}) is due to the fact that $\vec{h}_s$ is applied only on the surface.
A straightforward computation of Eq.~(\ref{chi4_def}) results in
\begin{multline}
  \chi_{4s} = \frac{1}{L^2}\Bigg[\<(\vec{M}_s\cdot \vec{M}_s)^2\> - \<\vec{M}_s\cdot \vec{M}_s\>^2\\
    - 2\sum_{i,j=1}^N\<M_s^{(i)} M_s^{(j)}\>\<M_s^{(i)} M_s^{(j)}\>\Bigg],
  \label{chi4_intermediate}
\end{multline}
where $M_s^{(i)}$ is the $i-$th component of $\vec{M}_s$.
The last term in Eq.~(\ref{chi4_intermediate}) is nonvanishing only when $i=j$.
Furthermore, with the O($N$) symmetry being unbroken, $\<(M_s^{(i)})^2\> = \<\vec{M}_s\cdot \vec{M}_s\>/N$.
Thus, Eq.~(\ref{chi4_intermediate}) simplifies to
\begin{equation}
  \chi_{4s} = \frac{1}{L^2}\left[\<(\vec{M}_s\cdot \vec{M}_s)^2\> - \left(\frac{N+2}{N}\right)\<\vec{M}_s\cdot \vec{M}_s\>^2\right].
  \label{chi4}
\end{equation}
The FSS behavior of $\chi_{4s}$ at the critical point is
\begin{equation}
  \chi_{4s} = A L^{6-4\dphi}\left(1 + \frac{C}{L}\right) + B,
  \label{FSSchi4}
\end{equation}
where we have anticipated that the leading scaling correction is $\propto L^{-1}$.
Indeed, in the improved lattice model considered here the leading irrelevant bulk scaling field is suppressed, and the next-to-leading correction, due to the lowest nonrotationally invariant irrelevant operator, decays fast as $\propto L^{-\omega_{\text{nr}}}$, with $\omega_{\text{nr}}\simeq 2$ \cite{CPRV-98}.
On the other hand, the surface operator spectrum contains a protected operator, the displacement operator, which encodes the broken translational invariance and whose existence is guaranteed on any conformal defect \cite{BGLM-16}; its dimension is $\Delta_D=3$, thus giving rise to corrections to scaling $\propto L^{-1}$.
The existence of such corrections was first pointed out in Ref.~\cite{CF-76} and can also be intuitively understood by an RG analysis of the scaling field associated with the size $L$ \cite{CPV-14}.
Although {\it a priori} the surface operator spectrum may contain more irrelevant perturbations, previous investigations on improved lattice models at the ordinary UC did not detect corrections decaying slower than $L^{-1}$ \cite{Hasenbusch-09b,Hasenbusch-11,PTTD-13}.
The results of this work support this picture, thus we conclude that Eq.~(\ref{FSSchi4}) reliably describes the leading scaling behavior of $\chi_{4s}$.

\begin{figure}
  \includegraphics[width=\linewidth]{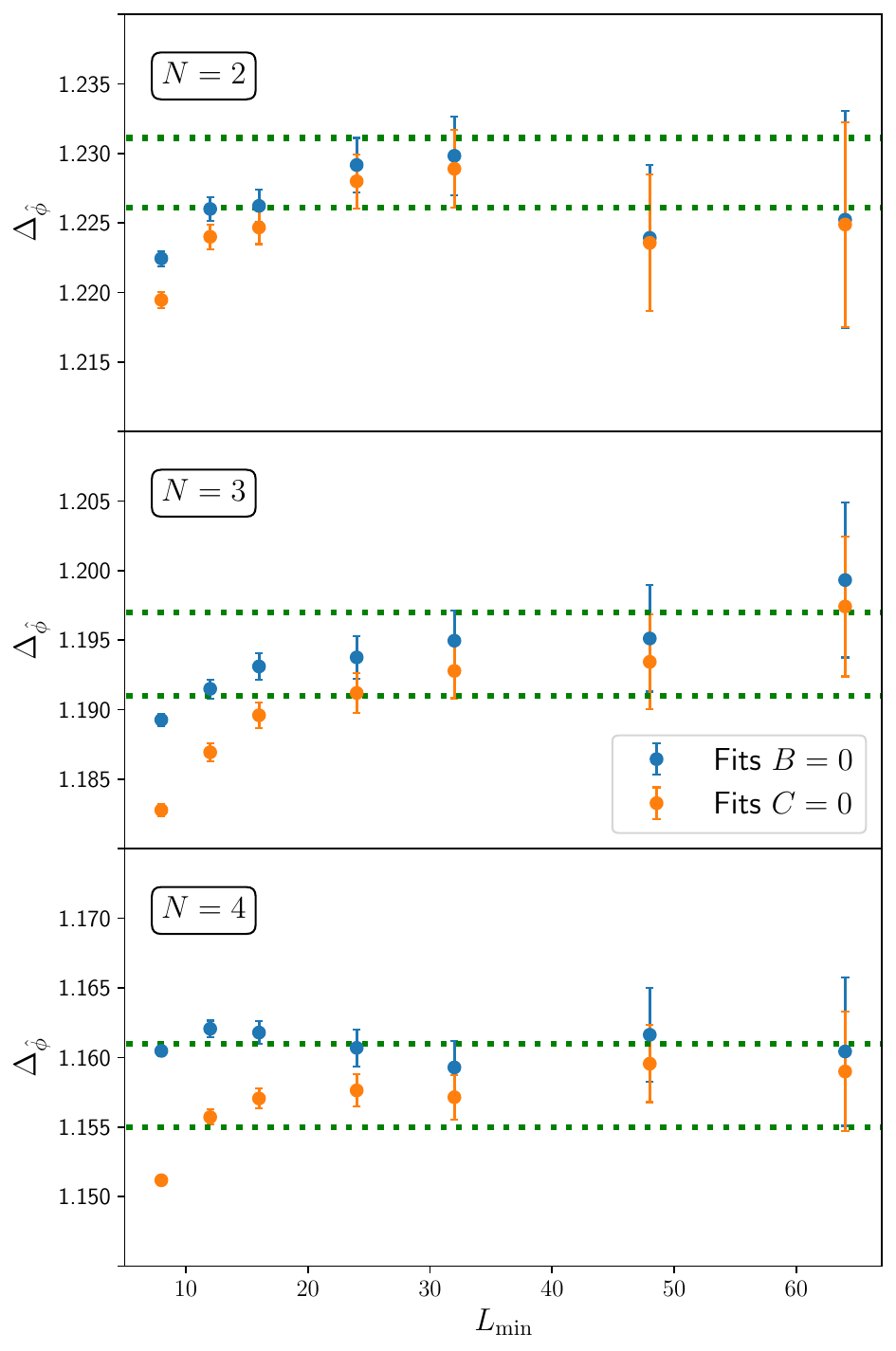}
  \caption{Fitted value of $\dphi$ for $N=2,3,4$, and as a function of the minimum lattice size $L_{\text{min}}$ taken into account. Results are obtained by fitting $\chi_{4s}$ to Eq.~(\ref{FSSchi4}), and fixing either $B=0$ or $C=0$.
    Dotted lines indicate an interval of one error bar around the final estimates given in Table \ref{table.results}.}
  \label{fits}
\end{figure}

To determine $\dphi$ we have sampled the model (\ref{model}) by means of high-precision MC simulations, for $N=2,3,4$ and lattice sizes $L=8,\ldots,384$.
As confirmed by fit results below, the leading exponent in Eq.~(\ref{FSSchi4}) $6-4\dphi \simeq 1.2 > 0$, so that $\chi_{4s}$ diverges and its FSS behavior is, unlike $\chi$, dominated by its singular part.
The background term $B$ represents a correction to scaling, effectively decaying as $L^{4\dphi-6} \sim L^{-1.2}$.
As this exponent is rather close to $1$, in the FSS analysis it is not technically feasible to reliably disentangle the two equally important sources of corrections $CL^{-1}$ and $BL^{4\dphi-6}$.
Therefore, we resolved to consider separately fits of MC data to Eq.~(\ref{FSSchi4}) by either fixing $B=0$ or $C=0$ \cite{SM}.
Such a procedure is expected to introduce a small bias in the fitted results, which nevertheless should be negligible for large enough lattice sizes: indeed, on increasing $L$, scaling corrections become numerically less significant, such that eventually both fits should give identical results.
Accordingly, and also in order to monitor residual subleading scaling corrections not considered in Eq.~(\ref{FSSchi4}), 
fits are repeated disregarding systematically the smallest lattices.
In Fig.~\ref{fits} we show fitted values of $\dphi$, as a function of the minimum lattice size $L_{\text{min}}$ taken into account.
Fits exhibit a good $\chi^2/{\text{d.o.f.}}$ (d.o.f. indicates the degrees of freedom) for $L_{\text{min}} \ge 16$ and in some cases also for $L_{\text{min}}=12$.
In line with the discussion above, we observe a small difference in the fitted value of $\dphi$ as obtained fixing either $B=0$ or $C=0$ in Eq.~(\ref{FSSchi4}).
Such a discrepancy is lifted when $L_{\text{min}}=16$ for $N=2$ and $L_{\text{min}}=32$ for $N=3,4$.
For $N=2$, on further discarding smaller lattices, we observe a slightly significant drift in the fitted values, hinting at residual scaling corrections: fits for $L_{\text{min}}=24, 32$ still give identical results when $B=0$ or $C=0$, but the fitted value of $\dphi$ is in marginal agreement with the results for $L_{\text{min}}=16$.
For this reason, we extract as a final estimate an average of the values obtained in the two fits for $L_{\text{min}}=24$, indicating a conservative error bar to be fully compatible with both fits; such an estimate also agrees with fit results for $L_{\text{min}}=16$ within one error bar.
For $N=3,4$, fits with $L_{\text{min}} \ge 32$ are perfectly stable and give indistinguishable results when setting either $B=0$ or $C=0$.
Accordingly, we quote as a final estimate of $\dphi$ an average of the values obtained in the two fits for $L_{\text{min}}=32$.
In Table \ref{table.results} we report our results, comparing them with previous determinations present in the literature.
We complement our estimate for $N=4$ by an analysis of the available field-theory series \cite{RG-80,RG-81,DD-81,*DD-81_erratum,Diehl-86,DS-94,DS-98,AS-95} and a TCB \cite{Gliozzi-13} calculation \cite{SM}.

\begin{table}
  \caption{Scaling dimension $\dphi$ of the leading surface operator at the ordinary transition of the three-dimensional O($N$) model. We compare present determinations with results obtained by the field-theoretic $\varepsilon-$expansion setting $\varepsilon=1$ ($\varepsilon-$exp), the massive field theory approach in $d=3$ (FT $d=3$) analyzed with a Pad\'e resummation, TCB, and previous MC simulations. The scaling dimension of the surface field $y_{h_1}$ is related to $\dphi$ by $y_{h_1}=2-\dphi$. The surface critical exponent $\eta_\parallel$ can be expressed in terms of $\dphi$ by $\dphi = (1+\eta_\parallel)/2$ \cite{Diehl-86}.}
  \begin{ruledtabular}
    \begin{tabular}{lcco}
      $N$ & Method & Ref. & \multicolumn{1}{c}{$\dphi$} \\
      \hline
          & $\varepsilon-$exp & \cite{RG-80,RG-81,DD-81,*DD-81_erratum,Diehl-86} & 1.19 \\
          & FT $d=3$ & \cite{DS-94,DS-98} & 1.211 \\
      $2$ & TCB      & \cite{GLMR-15}     & 1.2342(9) \\
          & MC       & \cite{DBN-05}      & 1.219(2) \\
          & MC       & This work          & 1.2286(25) \\
      \hline
          & $\varepsilon-$exp & \cite{RG-80,RG-81,DD-81,*DD-81_erratum,Diehl-86} & 1.153 \\
          & FT $d=3$ & \cite{DS-94,DS-98} & 1.169 \\
      $3$ & TCB      & \cite{GLMR-15}     & 1.198(1) \\
          & MC       & \cite{DBN-05}      & 1.187(2) \\
          & MC       & This work          & 1.194(3) \\
      \hline
          & $\varepsilon-$exp & \cite{RG-80,RG-81,DD-81,*DD-81_erratum,Diehl-86} & 1.125 \\
          & FT $d=3$ & This work      & 1.188 \\
      $4$ & TCB      & This work      & 1.172 \\
          & MC       & \cite{Deng-06} & 0.9798(12) \\
          & MC       & This work      & 1.158(3) \\
    \end{tabular}
  \end{ruledtabular}
  \label{table.results}
\end{table}

Another set of interesting universal quantities at a critical point are the fixed-point values of RG invariants.
Here, we consider two such observables: the ratio $(Z_a/Z_p)$ of the partition function with antiperiodic and periodic BC on a direction parallel to the surfaces, which can be efficiently sampled with the boundary-flip algorithm \cite{Hasenbusch-93,CHPRV-01}, and the combination $L\Upsilon$, where $\Upsilon$ is helicity modulus, i.e., the response to a torsion on a lateral direction \cite{FBJ-73}.
We notice that other commonly used RG invariants, such as the ratio $\xi/L$ of the surface correlations length $\xi$ over the size $L$, and the surface Binder cumulant $U_4=\<(\vec{M}_s\cdot \vec{M}_s)^2\>/\<\vec{M}_s\cdot \vec{M}_s\>^2$ are not particularly informative here: at the ordinary UC they acquire a trivial fixed-point value $(\xi/L)^*=0$ and $U_4^* = (N+2)/N$.
We fit RG-invariant observables $R$ to
\begin{equation}
  R = R^* \left(1 + A/L\right),
  \label{RGinvfit}
\end{equation}
leaving $R^*$ and $A$ as free parameters.
By judging conservatively the variation of the fit results on discarding smallest lattices, and the value of the $\chi^2/{\text{d.o.f.}}$, we obtain the estimates reported in Table \ref{table.RGinv} \cite{SM}.

\section{Discussion}
In this Letter we have studied the ordinary surface UC of the three-dimensional O($N$) model,
providing an accurate estimate of the scaling dimension $\dphi$ of the single relevant surface operator.
A comparison with previous MC estimates, reported in Table \ref{table.results}, indicates a small, but numerically significant, deviation from previous results, which are not compatible within one error bar.
Particularly significant is the difference between our estimate of $\dphi$ for $N=4$ and the result of Ref.~\cite{Deng-06}.
Here, we have simulated an improved model, where leading bulk scaling corrections are suppressed.
Furthermore, unlike previous studies which analyzed the surface susceptibility $\chi$, here $\dphi$ is extracted by a FSS analysis of the fourth cumulant $\chi_{4s}$; different than $\chi$, whose scaling behavior is dominated by its nonsingular part, $\chi_{4s}$ is divergent [compare Eq.~(\ref{FSSchis}) with Eq.~(\ref{FSSchi4})].
Hence, we expect our estimates to be more reliable than previous determinations, constituting a benchmark for future studies.
In Table \ref{table.results}, we also compare our results for $\dphi$ with TCB estimates.
This method introduces a small systematic error, whose magnitude is difficult to independently estimate \cite{GLMR-15,PKMGM-21}.
Still, TCB provides a rather good approximation of the boundary exponent, with a deviation of $\lesssim 1\%$ from the MC estimates.

In this work we have also studied the RG invariants at the ordinary transition.
These quantities are commonly used in the FSS analysis of second-order phase transitions.
In particular, they can be exploited in the scheme of FSS at fixed RG invariant \cite{Hasenbusch-99}, for which in some cases a significant reduction of error bars has been observed \cite{HPV-05,HPTPV-07,Wolff-09c,PT-11}; a comprehensive review of this method, together with a discussion of its implementation, can be found in Ref.~\cite{PT-21}.
Within this scheme,
our estimates of the fixed-point values of RG invariants reported in Table \ref{table.RGinv} provide a base for further numerical improvement of the critical exponents at the ordinary UC.

\begin{table}[t]
  \caption{Estimated critical value of RG invariants $(Z_a / Z_p)$ and $L\Upsilon$.}
  \begin{ruledtabular}
    \begin{tabular}{loo}
      $N$ & \multicolumn{1}{c}{$(Z_a / Z_p)^*$} & \multicolumn{1}{c}{$(L\Upsilon)^*$} \\
      \hline
      $2$ & 0.7016(2)  & 0.175(3) \\
      $3$ & 0.56480(8) & 0.1819(9) \\
      $4$ & 0.44588(9) & 0.1866(11) \\
    \end{tabular}
  \end{ruledtabular}
  \label{table.RGinv}
\end{table}

\section{Acknowledgments}
The author is grateful to Hans-Werner Diehl and Marco Meineri for useful communications.
F.P.T. is funded by the Deutsche Forschungsgemeinschaft (DFG, German Research Foundation), Project No. 414456783.
The author gratefully acknowledges the Gauss Centre for Supercomputing e.V. (www.gauss-centre.eu) for funding this project by providing computing time through the John von Neumann Institute for Computing (NIC) on the GCS Supercomputer JUWELS at Jülich Supercomputing Centre (JSC).
The author acknowledges support from the University of W\"urzburg, where this work was initiated.

\bibliography{francesco,extra}

\clearpage



\onecolumngrid
  \parbox[c][3em][t]{\textwidth}{\centering \large\bf Supplemental Material}
\smallskip
\twocolumngrid

\setcounter{equation}{0}
\renewcommand{\theHequation}{S.\arabic{equation}}
\renewcommand{\theequation}{S.\arabic{equation}}

\setcounter{figure}{0}
\renewcommand{\theHfigure}{S.\arabic{figure}}
\renewcommand{\thefigure}{S.\arabic{figure}}

\setcounter{table}{0}
\renewcommand{\theHtable}{S.\Roman{table}}
\renewcommand{\thetable}{S.\Roman{table}}

\section{Fits of the fourth cumulant $\chi_{4s}$}
\label{sm:fits}
\begin{table*}
  \caption{Fits of the fourth cumulant $\chi_{4s}$ to Eq.~(\ref{FSSchi4}), for $N=2,3,4$, and as a function of the minimum lattice size $L_{\text{min}}$ taken into account. Fit results are obtained fixing either $B=0$ or $C=0$.}
  \begin{ruledtabular}
    \begin{tabular}{ll.q.q}
          &                 &                \multicolumn{2}{c}{Fits $B=0$}                              &                 \multicolumn{2}{c}{Fits $C=0$} \\
      $N$ & $L_{\text{min}}$ & \multicolumn{1}{r}{$\dphi$} & \multicolumn{1}{r}{$\chi^2/{\text{d.o.f.}}$} & \multicolumn{1}{r}{$\dphi$} & \multicolumn{1}{c}{$\chi^2/{\text{d.o.f.}}$} \\
      \hline
      & $8$ & 1.22244(55) & 3.9& 1.21946(58) & 6.2 \\
      & $12$ & 1.22600(86) & 0.8& 1.22400(89) & 1.0 \\
      & $16$ & 1.2262(12) & 0.9& 1.2247(12) & 1.0 \\
  $2$ & $24$ & 1.2292(19) & 0.4& 1.2280(19) & 0.4 \\
      & $32$ & 1.2298(28) & 0.5& 1.2289(28) & 0.5 \\
      & $48$ & 1.2239(52) & 0.1& 1.2236(49) & 0.1 \\
      & $64$ & 1.2252(78) & 0.2& 1.2249(74) & 0.2 \\
      \hline
      & $8$ & 1.18927(43) & 3.1& 1.18278(44) & 11.1 \\
      & $12$ & 1.19150(68) & 1.2& 1.18694(66) & 3.3 \\
      & $16$ & 1.19311(94) & 0.5& 1.18960(90) & 0.9 \\
  $3$ & $24$ & 1.1938(15) & 0.6& 1.1912(14) & 0.7 \\
      & $32$ & 1.1950(22) & 0.5& 1.1928(20) & 0.6 \\
      & $48$ & 1.1951(38) & 0.7& 1.1934(34) & 0.7 \\
      & $64$ & 1.1993(56) & 0.6& 1.1974(50) & 0.6 \\
      \hline
      & $8$ & 1.16047(37) & 1.7& 1.15117(36) & 15.3 \\
      & $12$ & 1.16206(59) & 0.5& 1.15571(54) & 1.3 \\
      & $16$ & 1.16179(81) & 0.5& 1.15705(73) & 0.4 \\
  $4$ & $24$ & 1.1607(13) & 0.4& 1.1576(12) & 0.3 \\
      & $32$ & 1.1593(19) & 0.3& 1.1571(16) & 0.4 \\
      & $48$ & 1.1616(34) & 0.2& 1.1596(28) & 0.2 \\
      & $64$ & 1.1604(53) & 0.2& 1.1590(43) & 0.2 \\
    \end{tabular}
  \end{ruledtabular}
  \label{table.fits}
\end{table*}

In Table \ref{table.fits} we report results of fits of $\chi_{4s}$ to Eq.~(\ref{FSSchi4}), as a function of the minimum lattice size $L_{\text{min}}$ taken into account.

\section{Pad\'e resummation of massive field-theory results for $N=4$}
\label{sm:pade}
In this section we provide details of the analysis of the two-loop expansion of the boundary exponent $\eta_\parallel$ at the ordinary transition for $N=4$, obtained within the massive field-theory scheme at $d=3$.
Refs.~\cite{DS-94,DS-98} reports the following series expansion at two loops
\begin{equation}
  \begin{split}
  \eta_\parallel = &2 - \frac{N+2}{2(N+8)}\bar{u} - \frac{24(N+2)}{(N+8)^2}\left(C+\frac{N+14}{96}\right)\bar{u}^2 \\
  &+ O(\bar{u}^3),\qquad C=-0.105063,
  \label{etapar_massive}
  \end{split}
\end{equation}
where $\bar{u}$ is a rescaled renormalized coupling constant, such that the $\beta-$function is $\beta(\bar{u}) = \bar{u}-\bar{u}^2+O(\bar{u}^3)$.
The knowledge of the surface exponent $\eta_\parallel$ allows to determine the scaling dimension of the surface field by means of a standard scaling relation \cite{Diehl-86}
\begin{equation}
  \dphi=\frac{1+\eta_\parallel}{2}.
  \label{eta_parallel_dphi}
\end{equation}
In Ref.~\cite{DS-98}, the series is analyzed for $N=0,1,2,3$ using a Pad\'e resummation;
in particular, the diagonal $[1/1]$ Pad\'e approximant is argued to provide the best result.
In order to have a homogeneous comparison, also the value reported in Table \ref{table.results} for $N=4$ has been obtained with the $[1/1]$ Pad\'e approximant of Eq.~(\ref{etapar_massive}).
To carry out the calculation,
we have used the fixed-point value of $\bar{u}^*=1.369$ obtained in Ref.~\cite{AS-95} with a Pad\'e-Borel resummation of the six-loop expansion of $\beta(\bar{u})$.

\section{Conformal bootstrap determination of $\dphi$ for $N=4$}
\label{sm:cb}
In this section we use the truncated conformal bootstrap method \cite{Gliozzi-13} to compute the scaling dimension of the boundary field at the ordinary fixed point for $N=4$.
We recall in the following the essential ingredients of the method, following closely Ref.~\cite{GLMR-15}, which applied the technique to investigate the boundary exponents of the O($N$) model, for $N\le 3$.
We consider a three-dimensional conformal field theory in a semiinfinite space, where we introduce cartesian coordinates $x = ({\bf x}, z)$, such that the
the boundary surface is located at $z=0$.
We indicate bulk operators by $O$ and boundary ones by $\widehat{O}$, and their corresponding scaling dimensions $\Delta_O$ and $\Delta_{\widehat{O}}$.
Bulk primary operators satisfy an Operator Product Expansion (OPE)
\begin{equation}
  O_1(x)O_2(y) = \frac{\delta_{12}}{(x-y)^{2\Delta_{O_1}}} + \sum_k \lambda_{12k}C\left(x-y,\partial_y\right)O_k(y),
    \label{bulkOPE}
\end{equation}
where conformal invariance fixes the form of $C\left(x-y,\partial_y\right)$ and $\lambda_{12k}$ are the OPE coefficients.
A boundary OPE holds for $z\rightarrow 0$:
\begin{equation}
  O_1(x) = \frac{a_1}{(2z)^{\Delta_{O_1}}} + \sum_l \mu_{1l}D\left(z,\partial_z\right)\widehat{O}_l({\bf x}),
    \label{bdyOPE}
\end{equation}
where $D\left(z,\partial_z\right)$ is fixed by conformal invariance, and $\mu_{1l}$ are universal boundary OPE coefficients.
A crossing equation for the correlators $\< O_1(x) O_2(y)\>$ can be written employing either Eq.~(\ref{bulkOPE}) or Eq.~(\ref{bdyOPE}).
Introducing a Taylor expansion of the crossing equation, and truncating the OPE to $n_{\text{bulk}}$ bulk and $n_{\text{bdy}}$ boundary operators, one obtains \cite{GLMR-15}
\begin{equation}
  -\sum_{k=1}^{n_{\text{bulk}}} p_k f^k_{\text{bulk}}|_{\xi=1} + \sum_{l=1}^{n_{\text{bdy}}} q_l f^l_{\text{bdy}}|_{\xi=1} + a_1a_2 = \delta_{12}
\label{inhom}
\end{equation}
\begin{multline}
  -\sum_{k=1}^{n_{\text{bulk}}} p_k \partial_\xi^n f^k_{\text{bulk}}|_{\xi=1} + \sum_{l=1}^{n_{\text{bdy}}} q_l \partial_\xi^n\left[\xi^{(\Delta_{O_1}+\Delta_{O_2})/2}f^l_{\text{bdy}}\right]|_{\xi=1} \\
  + a_1a_2 \partial_\xi^n\xi^{(\Delta_{O_1}+\Delta_{O_2})/2}|_{\xi=1}= 0,\qquad n=1,\ldots,M,
\label{hom}
\end{multline}
where $M$ is a truncation parameter, $p_k=\lambda_{12k}a_k$, $q_l=\mu_{1l}\mu_{2l}$, and $f^k_{\text{bulk}}$, $f^l_{\text{bdy}}$ are the bulk and boundary conformal blocks;
in $d=3$ they are \cite{MO-95}
\begin{equation}
  \begin{split}
    f^k_{\text{bulk}} &=\xi^{\Delta_{O_k}/2} \cdot \\
    &{}_2F_1\left(\frac{\Delta_{O_k}+\Delta_{12}}{2},\frac{\Delta_{O_k}-\Delta_{12}}{2},\Delta_{O_k} - \frac{1}{2},-\xi\right),\\
  \Delta_{12} &\equiv \Delta_{O_1}-\Delta_{O_2},
  \end{split}
  \label{block_bulk}
\end{equation}
\begin{equation}
  f^l_{\text{bdy}} = \frac{1}{2\sqrt{\xi}}\left(\frac{4}{1+\xi}\right)^{\Delta_{\widehat{O}_l} - 1/2}\left(1+\sqrt{\frac{\xi}{1+\xi}}\right)^{-2(\Delta_{\widehat{O}_l} - 1)}.
  \label{block_bdy}
\end{equation}
As in Ref.~\cite{GLMR-15}, we set in Eqs.~(\ref{inhom})-(\ref{hom}) $O_1=O_2=\phi$ the leading O($4$) vector, i.e., the order parameter and
we truncate the OPE $\phi\times\phi$ [Eq.~(\ref{bulkOPE})] to the leading $n_{\text{bulk}}=2$ bulk scalars appearing on the right-hand side of Eq.~(\ref{bulkOPE}): the energy operator $\epsilon$ and the leading irrelevant scalar operator $\epsilon'$.
The leading operator appearing in the boundary OPE of $\phi$ is the leading surface vector $\hat{\phi}$.
We truncate the right-hand side of Eq.~(\ref{bdyOPE}) to $n_{\text{bdy}}=1$ boundary operators, i.e., to $\hat{\phi}$, whose scaling dimension $\dphi$ is the main target of the computation.
As input parameters for the computation,
we use the dimensions $\Delta_\epsilon$, $\Delta_{\epsilon'}$ and $\Delta_\phi$.
They are related to the usual exponents by $\Delta_\epsilon = 3-1/\nu$, $\Delta_{\epsilon'}=3+\omega$, $\Delta_\phi = (1+\eta)/2$.
We employ the recent MC results $\nu=0.74817(20)$, $\omega=0.755(5)$, $\eta=0.03624(8)$ \cite{Hasenbusch-21}.

At the ordinary transition the O($N$) symmetry is unbroken, hence $a_1=0$ in Eqs.~(\ref{bdyOPE}), (\ref{inhom}) and (\ref{hom}).
Thus, in the system of equations (\ref{inhom}) and (\ref{hom}) there are $4$ unknown parameters: $3$ OPE coefficients $p_\epsilon$, $p_{\epsilon'}$, $q_{\hat{\phi}}$ and $\dphi$.
In line with Ref.~\cite{PKMGM-21}, a truncated solution can be searched by setting $M=3$ in Eqs.~(\ref{inhom})-(\ref{hom}).
A nontrivial solution to the homogeneous linear set of equations (\ref{hom}) can exist only if the associated $3\times 3$ matrix is singular \cite{GLMR-15}.
To find a solution, we numerically search the minimum most singular value, as a function of the unknown dimension $\dphi$.
Within the numerical precision, we effectively find a zero of the most singular value, which corresponds to $\dphi\simeq 1.172$.
Given the general difficulty in estimating the systematic error due to the truncation \cite{GLMR-15,PKMGM-21}, we refrain here to quote an error bar.
The calculation allows also to determine the boundary OPE coefficient $\mu_\phi^2=0.867$.
We have checked that,
upon varying the input parameters within one error bar quoted in Ref.~\cite{Hasenbusch-21}, the resulting value of $\dphi$ changes on the fourth digit.
Hence, the uncertainty on the input parameters is negligible with respect to the systematic error due to the truncation.

\section{Fits of RG-invariant observables}
\label{sm:rginv}
\begin{figure}
  \includegraphics[width=\linewidth]{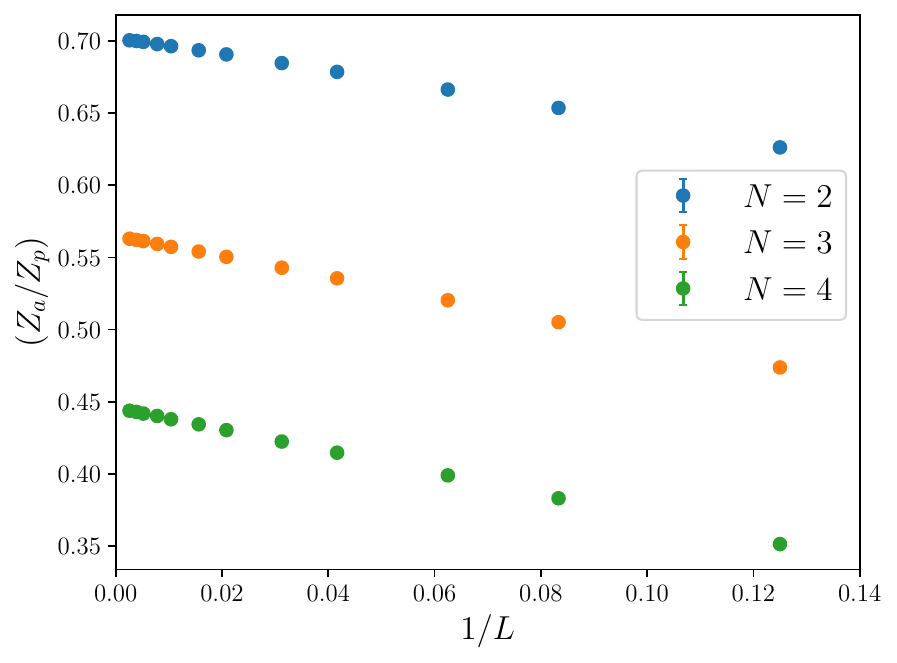}
  \caption{RG-invariant ratio $(Z_a / Z_p)$ as a function of $1/L$}
  \label{zazp}
\end{figure}
\begin{table}
  \caption{Fits of the RG invariant $(Z_a / Z_p)$ to Eq.~(\ref{RGinvfit}), for $N=2,3,4$, and as a function of the minimum lattice size $L_{\text{min}}$ taken into account.}
  \begin{ruledtabular}
    \begin{tabular}{ll.q}
      $N$ & $L_{\text{min}}$ & \multicolumn{1}{r}{$(Z_a / Z_p)^*$} & \multicolumn{1}{r}{$\chi^2/{\text{d.o.f.}}$} \\
      \hline
      & $8$ & 0.702585(30) & 74.7 \\ 
      & $12$ & 0.702325(32) & 11.9 \\
      & $16$ & 0.702242(33) & 5.8 \\
      & $24$ & 0.702176(37) & 3.9 \\
  $2$ & $32$ & 0.702129(40) & 3.2 \\
      & $48$ & 0.702072(49) & 3.1 \\
      & $64$ & 0.702051(59) & 3.7 \\
      & $96$ & 0.701997(77) & 4.6 \\
      & $128$ & 0.70184(10) & 4.5 \\
      & $192$ & 0.70156(14) & 0.7 \\
      \hline
      & $8$ & 0.565410(45) & 21.1 \\
      & $12$ & 0.565101(50) & 4.3 \\
      & $16$ & 0.565022(54) & 2.9 \\
      & $24$ & 0.564910(62) & 1.2 \\
  $3$ & $32$ & 0.564877(69) & 1.2 \\
      & $48$ & 0.564800(82) & 0.9 \\
      & $64$ & 0.564804(98) & 1.1 \\
      & $96$ & 0.56494(13) & 0.8 \\
      & $128$ & 0.56485(16) & 0.6 \\
      & $192$ & 0.56460(28) & 0.0 \\
      \hline
      & $8$ & 0.446078(51) & 2.6 \\
      & $12$ & 0.445969(59) & 1.4 \\
      & $16$ & 0.445901(67) & 1.0 \\
      & $24$ & 0.445862(78) & 1.0 \\
  $4$ & $32$ & 0.445856(92) & 1.2 \\
      & $48$ & 0.44577(11) & 1.1 \\
      & $64$ & 0.44572(13) & 1.2 \\
      & $96$ & 0.44589(18) & 1.0 \\
      & $128$ & 0.44566(25) & 0.5 \\
      & $192$ & 0.44589(42) & 0.6 \\
    \end{tabular}
  \end{ruledtabular}
  \label{table.fits_zazp}
\end{table}

In Fig.~\ref{zazp} we show the RG invariant $(Z_a / Z_p)$ for $N=2,3,4$.
As expected, it exhibits a linear behavior in $1/L$, confirming the Ansatz of Eq.~(\ref{RGinvfit}). In Table \ref{table.fits_zazp} we report fit results to Eq.~(\ref{RGinvfit}), from which we have extracted the estimates given in Table \ref{table.RGinv}.

The helicity modulus $\Upsilon$ is computed by inserting a torsion of an angle $\theta$ along one of the two directions parallel to the surfaces.
This can be obtained by replacing in the Hamiltonian (\ref{model})
\begin{equation}
  \vec{\phi}_{\vec{x}}\cdot\vec{\phi}_{\vec{x}+\hat{e}_1} \rightarrow \vec{\phi}_{\vec{x}} R_{\alpha,\beta}(\theta)\vec{\phi}_{\vec{x}+\hat{e}_1}, \quad \vec{x}=(x_1=x_{1,f},x_2,x_3),
  \label{torsion}
\end{equation}
where $R_{\alpha,\beta}(\theta)$ is a $2\times 2$ matrix that rotates the $\alpha$ and $\beta$ components of $\vec{\phi}$ by an angle $\theta$.
With a slight change of notation,
in Eq.~(\ref{torsion}) $\vec{x}=(x_1,x_2,x_3)$ indicates the lattice site as a three-dimensional vector, and $\hat{e}_1$ is the unit vector in the $1-$direction; within this notation, the surfaces are located at $x_3=1,L$.
The torsion of Eq.~(\ref{torsion}) is placed on a ``defect'' plane $x_1=x_f$ and acts on the direction $1$.
With a suitable change of variables, it is possible to ``smear out'' the torsion over the entire length orthogonal to the defect plane \cite{PT-20}.
The helicity modulus $\Upsilon$ is defined as \cite{FBJ-73}
\begin{equation}
  \Upsilon \equiv \frac{L}{S} \frac{\partial^2 F(\theta)}{\partial \theta^2}\Big|_{\theta=0},
  \label{hslicity_def}
\end{equation}
where $S=L_\parallel^2=L^2$ is the area of the surfaces, and $F$ is the free energy in units of $k_BT$.

As in Ref.~\cite{PT-20}, an improved estimator for $\Upsilon$ can be obtained by averaging over the $N(N-1)/2$ pairs of components $(\alpha,\beta)$ where the torsion is applied, as well as on the two possible lateral directions where the torsion is inserted.
One obtains the expression:
\begin{equation}
  \begin{split}
    \Upsilon &= \frac{1}{2L^3}\left[\frac{2}{N}\langle E\rangle - \sum_{\hat{e}=\hat{e}_1,\hat{e}_2}\frac{2}{N(N-1)}\sum_{\alpha<\beta} \langle \left(T^{(\alpha,\beta)}_{\hat{e}}\right)^2\rangle\right],\\
    E &\equiv \beta \sum_{\substack{\vec{x} \in {\rm bulk} \\ \hat{e}=\hat{e}_1,\hat{e}_2}}\vec{\phi}_{\vec{x}}\cdot\vec{\phi}_{\vec{x}+\hat{e}} + \beta_s\sum_{\substack{\vec{x} \in {\rm surface} \\ \hat{e}=\hat{e}_1,\hat{e}_2}}\vec{\phi}_{\vec{x}}\cdot\vec{\phi}_{\vec{x}+\hat{e}}\\
    T^{(\alpha,\beta)}_{\hat{e}} &\equiv \beta\sum_{\vec{x} \in {\rm bulk}}\left(\phi_{\vec{x}}^{(\alpha)}\phi_{\vec{x}+\hat{e}}^{(\beta)} - \phi_{\vec{x}}^{(\beta)}\phi_{\vec{x}+\hat{e}}^{(\alpha)}\right)\\
    &+ \beta_s\sum_{\vec{x} \in {\rm surface}}\left(\phi_{\vec{x}}^{(\alpha)}\phi_{\vec{x}+\hat{e}}^{(\beta)} - \phi_{\vec{x}}^{(\beta)}\phi_{\vec{x}+\hat{e}}^{(\alpha)}\right),\\
  \end{split}
  \label{Upsilon_imp}
\end{equation}
which is a generalization of the formula reported in Ref.~\cite{PT-20}.

\begin{figure}
  \includegraphics[width=\linewidth]{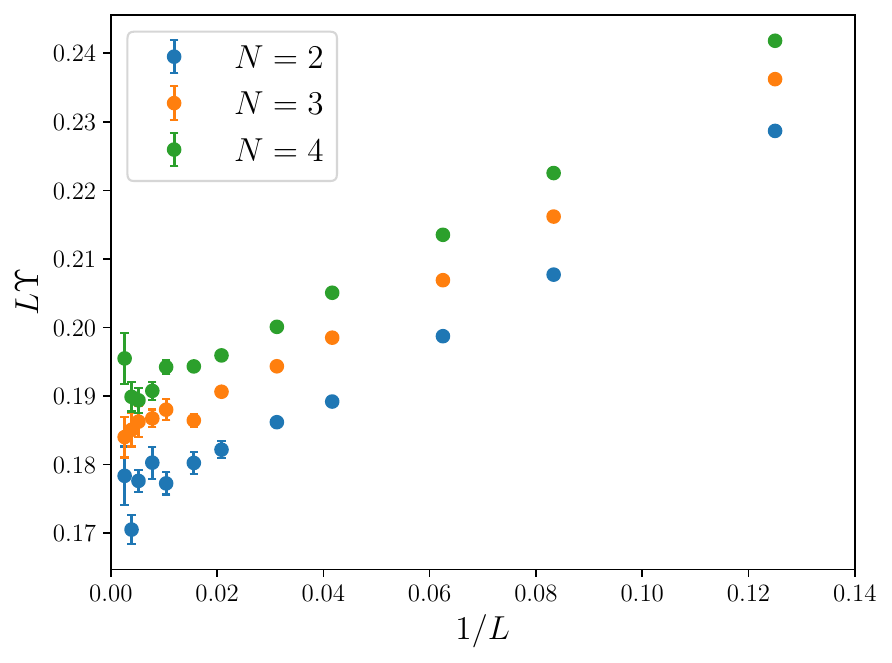}
  \caption{RG invariant $L\Upsilon$ as a function of $1/L$}
  \label{LY}
\end{figure}
\begin{table}
  \caption{Fits of the RG invariant $L\Upsilon$ to Eq.~(\ref{RGinvfit}), for $N=2,3,4$, and as a function of the minimum lattice size $L_{\text{min}}$ taken into account. Here we consider only MC data for $L\le L_{\text{max}} = 48$ (see text).}
  \begin{ruledtabular}
    \begin{tabular}{ll.q}
      $N$ & $L_{\text{min}}$ & \multicolumn{1}{r}{$(L\Upsilon)^*$} & \multicolumn{1}{r}{$\chi^2/{\text{d.o.f.}}$} \\
      \hline
      & $8$ & 0.16922(43) & 7.7 \\
  $2$ & $12$ & 0.17222(72) & 1.1 \\
      & $16$ & 0.1727(10) & 1.5 \\
      & $24$ & 0.1759(22) & 0.2 \\
      \hline
      & $8$ & 0.17861(26) & 13.8 \\
  $3$ & $12$ & 0.18103(43) & 1.9 \\
      & $16$ & 0.18207(63) & 0.2 \\
      & $24$ & 0.1825(12) & 0.1 \\
      \hline
      & $8$ & 0.18564(18) & 10.1 \\
  $4$ & $12$ & 0.18704(29) & 0.9 \\
      & $16$ & 0.18720(43) & 1.2 \\
      & $24$ & 0.18637(82) & 1.0 \\
    \end{tabular}
  \end{ruledtabular}
  \label{table.fits_LY_48}
\end{table}
\begin{table}
  \caption{Same as Table \ref{table.fits_LY_48} for $L_{\text{max}}=64$.}
  \begin{ruledtabular}
    \begin{tabular}{ll.q}
      $N$ & $L_{\text{min}}$ & \multicolumn{1}{r}{$(L\Upsilon)^*$} & \multicolumn{1}{r}{$\chi^2/{\text{d.o.f.}}$} \\
      \hline
      & $8$ & 0.16944(42) & 7.1 \\
      & $12$ & 0.17242(67) & 1.0 \\
  $2$ & $16$ & 0.17299(94) & 1.1 \\
      & $24$ & 0.1754(17) & 0.2 \\
      & $32$ & 0.1743(26) & 0.0 \\
      \hline
      & $8$ & 0.17865(26) & 11.1 \\
      & $12$ & 0.18086(41) & 1.7 \\
  $3$ & $16$ & 0.18159(56) & 1.1 \\
      & $24$ & 0.18130(96) & 1.6 \\
      & $32$ & 0.1806(15) & 3.0 \\
      \hline
      & $8$ & 0.18575(18) & 9.3 \\
      & $12$ & 0.18713(28) & 0.9 \\
  $4$ & $16$ & 0.18733(39) & 1.0 \\
      & $24$ & 0.18693(67) & 1.2 \\
      & $32$ & 0.1881(10) & 0.3 \\
     \end{tabular}
  \end{ruledtabular}
  \label{table.fits_LY_64}
\end{table}
In Fig.~\ref{LY} we show the RG-invariant combination $L\Upsilon$ as a function of $1/L$.
While we observe a linear behavior in $1/L$ for $L\lesssim 64$, MC data for larger value of $L$ are affected by considerable noise.
Indeed, the formula for $\Upsilon$ (\ref{Upsilon_imp}) involves a delicate subtraction: the first term is the energy density, and converges to a finite positive value for $L\rightarrow\infty$. The second term is manifestly positive and, subtracted to the energy term, must gives a quantity $O(1/L)$, such that $L\Upsilon$ acquires a finite nontrivial value for $L\rightarrow\infty$.
These considerations suggest a numerical instability of the sampled observable, which is clearly visible in Fig.~\ref{LY}.
Given these technical difficulties, we have decided to use only the MC data for $L\le L_{\text{max}}=48, 64$ in the analysis.
Corresponding fit results are reported in Tables \ref{table.fits_LY_48}, \ref{table.fits_LY_64}.
From the fits obtained with $L_{\text{max}}=48$, judging conservatively the stability of the results and the value of $\chi^2/{\text{d.o.f.}}$, we can estimate $(L\Upsilon)^*=0.176(2)$ for $N=2$, $(L\Upsilon)^*=0.1821(6)$ for $N=3$, and $(L\Upsilon)^*=0.1864(9)$ for $N=4$.
Using the the fits for $L_{\text{max}}=64$, we obtain $(L\Upsilon)^*=0.174(3)$ for $N=2$, and $(L\Upsilon)^*=0.1873(4)$ for $N=4$; for $N=3$ the fit with $L_{\text{min}}=16$ gives $(L\Upsilon)^* = 0.18159(56)$, although with a slightly large $\chi^2/{\text{d.o.f.}}=1.1$, while other fits have a large $\chi^2/{\text{d.o.f.}}$.
The final values quoted in Table \ref{table.RGinv} are an average of the estimates found with $L_{\text{max}}=48, 64$, with an uncertainty fixed so as to have fully compatibility to such estimates within one error bar.

\end{document}